\documentclass[aps,pre,showpacs,noshowkeys,amsmath,amssymb,amsfonts,superscriptaddress,longbibliography,reprint]{revtex4-2}
\usepackage[english]{babel}

\usepackage{graphicx}
\usepackage{bm}
\usepackage{physics}
\usepackage{mathtools}
\usepackage{gensymb}
\usepackage{booktabs}

\newcommand{\Lc}{\ensuremath{L_\text{c}}}
\newcommand{\Ta}{\ensuremath{T_\text{a}}}
\newcommand{\Req}{\ensuremath{R_\text{eq}}}
\newcommand{\Leq}{\ensuremath{L_\text{eq}}}
\newcommand{\teq}{\ensuremath{\theta_\text{eq}}}

\usepackage{comment}
\usepackage{siunitx}

\usepackage{caption}
\usepackage{subcaption}
\DeclareCaptionLabelSeparator{bar}{~\rule[-0.4ex]{0.2ex}{1em}~}
\DeclareCaptionLabelFormat{subfor}{\textbf{#2}}
\newcommand*\bfcaption[2]{\caption[#1]{\textbf{#1.}#2}}
\usepackage[dvipsnames]{xcolor}
\definecolor{UBcolor}{HTML}{007CC1}

\usepackage[colorlinks=true,pdfnewwindow=true,linkcolor=UBcolor,citecolor=UBcolor,urlcolor=UBcolor,breaklinks=true,linktocpage]{hyperref}
\usepackage[all]{hypcap}
\usepackage[nameinlink,capitalise]{cleveref}
\AddToHook{cmd/appendix/before}{%
  \crefalias{section}{appendix}%
  \crefalias{subsection}{appendix}%
}
\usepackage{xr}
\crefname{appendix}{Appendix}{Appendices}
\Crefname{appendix}{Appendix}{Appendices}
\begin{document}

\title{Active wetting transition in cell aggregates}

\author{MJ Franco O\~{n}ate}
\email{mj.franco-onate@ub.edu}
\affiliation{Departament de F\'{i}sica de la Mat\`{e}ria Condensada, Facultat de F\'{i}sica, Universitat de Barcelona, Barcelona, Spain}
\affiliation{Universitat de Barcelona Institute of Complex Systems (UBICS), Barcelona, Spain}

\author{Hanno I. Hennighausen}
\affiliation{Max Planck Institute for the Physics of Complex Systems, Dresden, Germany}
\affiliation{Center for Systems Biology Dresden, Dresden, Germany}

\author{Ricard Alert}
\email{ricard.alert@ub.edu}
\affiliation{Departament de F\'{i}sica de la Mat\`{e}ria Condensada, Facultat de F\'{i}sica, Universitat de Barcelona, Barcelona, Spain}
\affiliation{Universitat de Barcelona Institute of Complex Systems (UBICS), Barcelona, Spain}
\affiliation{Max Planck Institute for the Physics of Complex Systems, Dresden, Germany}
\affiliation{Center for Systems Biology Dresden, Dresden, Germany}
\affiliation{Cluster of Excellence Physics of Life, TU Dresden, Dresden, Germany}
\affiliation{Instituci\'{o} Catalana de Recerca i Estudis Avan\c{c}ats (ICREA), Barcelona, Spain}

\begin{abstract}
During many biological processes, cell aggregates spread on substrates in a process of active wetting. The existing models of tissue spreading, however, do not explicitly capture the state of partial wetting, in which the cell aggregate reaches a non-zero contact angle. Here, modeling cell aggregates as active polar liquid droplets, we predict a transition between partial and complete wetting. This transition emerges from the competition between tissue surface tension and active traction forces, which is captured in an active Young-Dupr\'{e} equation. The active wetting transition is discontinuous, and it features a region of parameters in which both partial and complete wetting are possible. Unlike in passive wetting, the transition depends on dynamical parameters such as the tissue viscosity and substrate friction. Finally, adding contractile intercellular active stresses leads to higher contact angles and even full dewetting. Overall, our work predicts an active wetting transition for tissues and it captures the partial wetting states often seen in experiments.
\end{abstract}

\maketitle

The spreading of epithelial tissues plays a central role in fundamental biological processes such as wound healing, embryonic development, and cancer progression \cite{Friedl2009,Ladoux2017,Cheung2025,Behrndt2012,Morita2017,Wallmeyer2018,Bondarenko2023,Huycke2024,Blauth2024,Wu2026,Wimalasena2026,Cavanaugh2026}. Over a decade ago, by examining how three-dimensional cell aggregates spread on a substrate, several works proposed an analogy with the wetting of liquid droplets \cite{Douezan2011,Douezan2012c,Douezan2012a,Gonzalez-Rodriguez2012,Beaune2014,Yousafzai2022a,Wang2023a,Pahlavan2026}. Although it rationalizes some experimental observations, the analogy to passive wetting overlooks the inherently active nature of living tissues, where active cellular forces drive collective cell migration and tissue shape changes. To address this challenge, subsequent work developed a model of tissue spreading as a process of active wetting \cite{Perez-Gonzalez2019,Alert2020,Alert2021d}. This model explained the wetting and dewetting dynamics of epithelial cell layers \cite{Perez-Gonzalez2019}, as well as fingering instabilities \cite{Perez-Gonzalez2019,Alert2019,Trenado2021} and processes of directed migration such as tissue durotaxis \cite{Alert2019a,Pi-Jauma2022,Pallares2023}.

So far, however, the active wetting model has not captured the transition between states of complete and partial wetting, which are often observed experimentally \cite{Douezan2011,Douezan2012c,Ravasio2015,Pallares2023,Conti2024,Aslemarz2024,Lemahieu2025a}. Wetting transitions have been studied in models of other active systems \cite{Thiele2026}, such as the actin cortex \cite{Joanny2013}, biomolecular condensates \cite{Zhao2021b,Zhao2024e,Liese2025}, bacterial biofilms \cite{Trinschek2017}, active nematics \cite{Joanny2012,Adkins2022,Coelho2023,Li2025,Chandel2025,Mandal2026}, and self-propelled particles \cite{Wittmann2016,Smeets2016,Sepulveda2017,Sepulveda2018,Wysocki2020,Das2020,Turci2021a,Turci2024,Mangeat2024,FinsCarreira2024,Das2025,Zhao2026,Matsuzawa2026,Grodzinski2026,Grodzinski2026a}. All these cases present important differences with the spreading of cell aggregates. In the case of self-propelled particles, for example, the droplet arises through motility-induced phase separation, and it is made cohesive by inward-pointing particles at the droplet boundary \cite{Fily2012,Bialke2013,Speck2020,Zhang2021i}. In contrast, in tissues, cohesion arises from cell-cell adhesion \cite{Lecuit2015,Ladoux2017}, and cells at the tissue edge polarize outwards \cite{Ladoux2016,Alert2020,Smeets2016,Panigrahi2025}. Thus, a theoretical description of partial active wetting in tissues remains lacking.

Here, we report a transition between partial and complete wetting in a model of cell aggregates as active droplets. Partial wetting arises from the balance of active traction forces, which promote spreading, and tissue surface tension, which opposes it. We obtain an active Young-Dupr\'{e} condition that captures this competition and determines the steady-state contact angle of the tissue. These results could explain experimental observations of increased tissue wettability on stiffer or more adhesive substrates \cite{Douezan2012c,Ravasio2015,Pallares2023,Conti2024,Lemahieu2025a}, where cells exert stronger tractions \cite{Perez-Gonzalez2019,Pallares2023}. Complete wetting appears when tractions overcome surface tension. We find that the wetting transition is discontinuous and features a region of bistability, in which the tissue can either partially or completely wet the surface depending on tissue size. Moreover, the wetting transition is modulated by dynamical parameters such as tissue viscosity and the cell-substrate friction coefficient. The fact that the wetting steady state depends on tissue size and dynamical parameters makes the active wetting transition fundamentally different from its passive counterpart, for which the equilibrium contact angle is set solely by surface tensions. These dependencies also provide additional knobs to control tissue wetting, for example via changes in tissue viscosity, which were recently observed in cancer cell aggregates \cite{Marchesi2026}, zebrafish \cite{Naik2026}, and mouse embryos \cite{Cavanaugh2026}.

\bigskip

\begin{figure}[tb!]
    \centering
    \includegraphics[width=\linewidth]{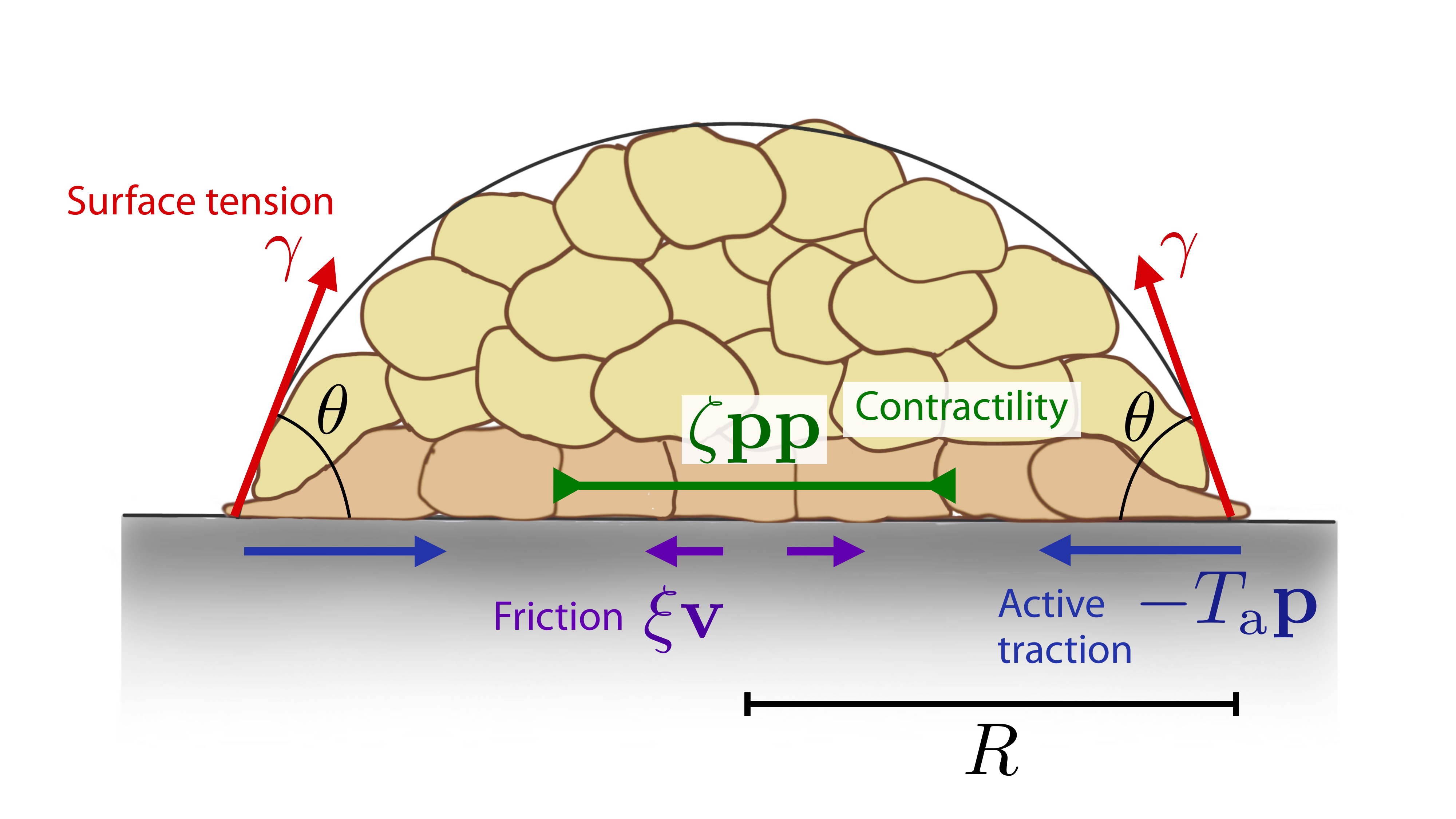}
    \bfcaption{Schematic of the model}{ The basal cell layer, which drives tissue spreading, is displayed in darker color.}
    \label{fig:schematic_model}
\end{figure}

\noindent\textbf{Active wetting model.} Building on Refs.~\cite{Perez-Gonzalez2019, Pallares2023}, we describe tissue spreading as the wetting of an active droplet. The droplet takes the shape of a spherical cap with fixed volume $\mathcal{V}$ and varying contact radius $R(t)$ and contact angle $\theta(t)$ (\cref{fig:schematic_model}). We consider that tissue spreading is driven by the circular basal layer of cells (darker color in \cref{fig:schematic_model}). These cells are in contact with the substrate and exhibit front-back polarity, which we describe through a polarity field $\mathbf{p}(\mathbf{r},t)$ \cite{Alert2020}. Following Ref.~\cite{Perez-Gonzalez2019}, we assume that cells in the basal layer are unpolarized in the center and polarized radially outwards at the edge ($r=R$), where we impose $p_r(R,t) = 1$, with $r$ the radial coordinate. Based on the physics of polar liquid crystals \cite{deGennes1993}, we assign a free energy
\begin{equation} \label{eq free-energy}
F = \int \left[ \frac{a}{2} \mathbf{p}^2 + \frac{K}{2} \left|\bm{\nabla}\mathbf{p}\right|^2 \right] \dd^2\mathbf{r}
\end{equation}
to the polarity field. This free energy favors the unpolarized state $\mathbf{p}=0$ with a restoring coefficient $a>0$ and penalizes spatial variations of the polarity with a Frank constant $K$. Assuming that the polarity field relaxes much faster than the time scale of tissue flows \cite{Perez-Gonzalez2019}, it achieves the equilibrium state corresponding to the free energy minimum, which yields
\begin{equation} \label{eq polarity}
\Lc^2 \nabla^2 \mathbf{p} = \mathbf{p},
\end{equation}
where $\Lc = \sqrt{K/a}$ is the length over which the polarity decays from its maximum at the edge to zero in the center.

We next establish force balance in the basal cell layer. Neglecting inertial forces since cell flows occur at very low Reynolds numbers, force balance reads
\begin{equation} \label{eq force_balance}
\bm{\nabla}\cdot \bm{\sigma} = \xi \mathbf{v} - \Ta \mathbf{p}.
\end{equation}
The right-hand side corresponds to the traction forces exchanged between the cells and the substrate, which include (i) friction with coefficient $\xi$ and proportional to the two-dimensional tissue flow field $\mathbf{v}(\mathbf{r},t)$, and (ii) active traction forces of magnitude $\Ta$ exerted by polarized cells as described by the polarity field $\mathbf{p}$ (\cref{fig:schematic_model}). The left-hand side corresponds to the internal force density, given by the divergence of the stress tensor $\bm{\sigma}$. The internal stresses, given by
\begin{equation} \label{eq stress}
\bm{\sigma} = \eta \left(\bm{\nabla} \mathbf{v} + \left(\bm{\nabla} \mathbf{v}\right)^T \right) - \zeta \mathbf{p} \mathbf{p},
\end{equation}
include (i) viscous stresses with viscosity $\eta$ that capture internal dissipation, for example due to cell-cell friction during tissue flows, and (ii) cell-cell active stresses with coefficient $\zeta$, which we take to be contractile by setting $\zeta<0$. We refer to this last term simply as contractility (\cref{fig:schematic_model}). Note that the stress has a non-zero trace. Thus, the basal cell layer is compressible, which allows it to change area as the tissue spreads or retracts. These area changes can be accompanied by cell exchanges between the basal and the upper layers, such that the total volume of the cell aggregate remains constant.

Finally, we specify a boundary condition for the tissue flow field $\mathbf{v}$. Following Ref.~\cite{Pallares2023}, and based on the physics of wetting \cite{DeGennes1985,DeGennes2004,Bonn2009}, we impose a generalized Young--Dupr\'{e} condition at the contact line:
\begin{equation}
    \sigma_{rr}(R) = -\gamma \cos\theta,
    \label{eq:Young-Dupre}
\end{equation}
where $\theta$ is the contact angle (\cref{fig:schematic_model}). This equation balances the horizontal component of the tissue surface tension $\gamma$ with the radial tension $\sigma_{rr}$ of the basal layer at the tissue edge.

\bigskip

\noindent\textbf{Active partial--to--complete wetting transition.} First, we focus on the simplest realization of the model by ignoring friction and contractility. The model has two sources of dissipation: viscosity and friction. They define the screening length $\lambda = \sqrt{\eta/\xi}$, over which the flow extends before being damped by substrate friction. For epithelial monolayers, typical values of the screening length are $\lambda \sim 0.2 - 0.6$ mm \cite{Perez-Gonzalez2019,Heinrich2020}. For small tissues, with contact radius $R< \lambda$, viscosity dominates, and friction has a small effect on the tissue flow \cite{Alert2019}. We focus on this regime in this section by setting $\xi=0$. We also set the contractility $\zeta = 0$ to isolate the competition between the two key forces in our wetting transition: active traction $\Ta$ and surface tension $\gamma$.

With these simplifications, and assuming radial symmetry, such that $\mathbf{p} = p(r) \hat{\mathbf{r}}$ and $\mathbf{v} = v(r) \hat{\mathbf{r}}$, we solve the model analytically. Solving \cref{eq polarity} in polar coordinates with the boundary condition $p(R)=1$, we obtain $p(r) = I_1(r/\Lc)/I_1(R/\Lc)$, with $I_n$ being the modified Bessel function of the first kind and order $n$. Introducing this solution into \cref{eq force_balance,eq stress}, we solve the force balance equation with the boundary condition \cref{eq:Young-Dupre} and the symmetry condition $v(0)=0$ to obtain the flow field $v(r)$ (\cref{solutions}). From this solution, we obtain the spreading velocity $V = v(R)$, which reads
\begin{equation}
    V= \frac{R}{\eta} \left( \Ta \Lc \frac{I_2(R/\Lc)}{I_1(R/\Lc)}
    - \gamma\cos{\theta}\right).
    \label{eq:edge_velocity}
\end{equation}
Note that the contact radius $R$ and the contact angle $\theta$ are not independent; given the volume of the cell aggregate, $\mathcal{V} = \pi R^3 f(\theta)/3$ with $f(\theta)= \left( 2 - 3\cos\theta + \cos^3\theta \right)/\sin^3\theta$, each value of $R$ corresponds to a given value of $\theta$. In the following, we analyze the spreading velocity \cref{eq:edge_velocity} as a function of the contact radius $R$.

\begin{figure}[tb!]
    \centering
\includegraphics[width=\columnwidth]{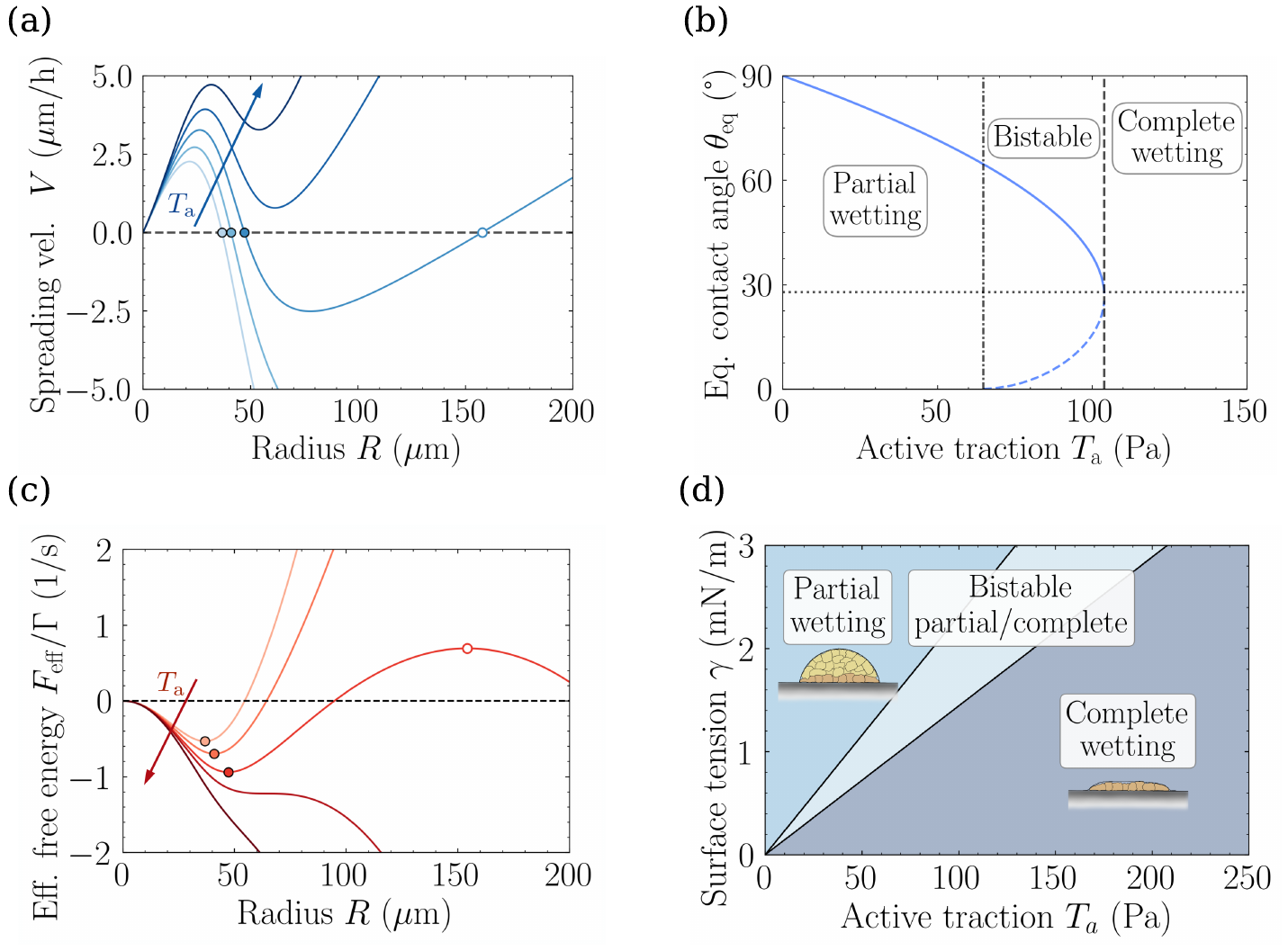}
  {\phantomsubcaption\label{Fig VR_traction}}
  {\phantomsubcaption\label{Fig theta_traction}}
  {\phantomsubcaption\label{Fig effective_free_energy}}
  {\phantomsubcaption\label{Fig phase_diagram}}
\bfcaption{Active partial--to--complete wetting transition}{ (a) Spreading velocity as a function of the contact radius $R$ for increasing active traction values $\Ta = 5, 41, 78, 114, 150$ Pa. Filled circles mark the stable equilibrium radius, corresponding to the partial wetting state. Empty circles mark unstable fixed points, above which there is complete wetting. (b) Stable (solid) and unstable (dashed) equilibrium contact angle $\teq$ as a function of $\Ta$. The stable and unstable branches meet at a saddle-node bifurcation at $\Ta^*$. Above this value, the contact angle jumps discontinuously from $\theta^*$ in the partial wetting regime to $\theta = 0$ in the complete wetting regime. (c) Effective free energy of the system for increasing values of $\Ta = 5, 41, 78, 104, 150$ Pa. The free energy develops an inflection point at the transition $\Ta^*= 104$ Pa. Below it, the free energy minimum determines the stable radius for partial wetting. (d) Phase diagram of active tissue wetting. Other parameter values are listed in \cref{tab_app: parameters_2D}.}
    \label{fig:partial_to_total}
\end{figure}

For small radii, for which $\theta > 90^\circ$, both active traction and surface tension pull the contact line outward. Hence, the tissue spreads with $V>0$ (\cref{Fig VR_traction}). However, when the contact angle drops below $90^\circ$, surface tension pulls backwards, thus opposing active tractions and slowing down spreading. Eventually, the tissue reaches an equilibrium contact radius at which spreading stops: $V(\Req)=0$ (filled circles in \cref{Fig VR_traction}). This equilibrium state is a stable fixed point of the spreading dynamics, and it has a non-zero contact angle: it corresponds to a state of partial wetting.

From \cref{eq:edge_velocity}, the condition $V(\Req)=0$ implies
\begin{equation}
    \Ta \Lc \frac{I_2(\Req/\Lc)}{I_1(\Req/\Lc)} = \gamma \cos{\teq}.
    \label{eq:dynamic_equilibrium}
\end{equation}
This equation is an active analogue of the classic Young-Dupr\'{e} condition, $\gamma_\text{SG} - \gamma_\text{SL} = \gamma_\text{LG} \cos\theta$, which sets the equilibrium contact angle in passive droplets as a balance of surface tensions between the solid, liquid, and gas phases \cite{DeGennes1985,DeGennes2004,Bonn2009}. Analogously, \cref{eq:dynamic_equilibrium} encodes the balance between the outward-pointing in-plane tension due to active tractions $\sim \Ta \Lc$ and the inward-pointing surface tension $\gamma$. Unlike for passive wetting, \cref{eq:dynamic_equilibrium} encodes a state of mechanical, but not thermodynamic, equilibrium. Moreover, the active term on the left-hand side depends on the contact radius. For large radii, $R\gg \Lc$, which is often the case in experiments \cite{Perez-Gonzalez2019,Heinrich2020}, \cref{eq:dynamic_equilibrium} reduces to
\begin{equation}
\Ta \Lc \approx \gamma \cos{\teq},
\end{equation}
where we can directly identify the active tension $\Ta \Lc$, which plays the role of $\gamma_\text{SG} - \gamma_\text{SL}$ in passive wetting.

As active traction increases, the equilibrium contact radius increases (\cref{Fig VR_traction}), and hence the contact angle decreases (\cref{Fig theta_traction}). Above a certain traction $T_\text{min}$, the curve $V(R)$ no longer has one zero but two (\cref{Fig VR_traction}). The second zero takes place at a critical radius $R_\text{u}$, which corresponds to an unstable fixed point (empty circle in \cref{Fig VR_traction}). It separates two regimes: Larger tissues ($R>R_\text{u}$) spread fully, reaching a state of complete wetting where $\theta = 0$, whereas smaller tissues ($R<R_\text{u}$) evolve towards the stable partial wetting state at $\Req$. Therefore, in this parameter regime, the system exhibits size-controlled bistability between partial and complete wetting (\cref{Fig theta_traction}). Finally, for even higher active traction, above $\Ta^*$, the curve $V(R)$ has no zeros (\cref{Fig VR_traction}); the tissue spreads fully, and only the complete wetting state exists (\cref{Fig theta_traction}). \Cref{Fig phase_diagram} summarizes these results in a phase diagram of active wetting.

\bigskip

\noindent\textbf{Nature of the wetting transition.} At the transition point $\Ta^*$, the stable and unstable fixed points of the spreading dynamics $\dot{R} = V(R)$ meet at a saddle-node bifurcation (\cref{Fig theta_traction}). When crossing the transition at $\Ta^*$, the equilibrium contact angle $\teq$ jumps discontinuously from a non-zero value $\theta^*$ to $\theta=0$ (\cref{Fig theta_traction}). This discontinuous behavior is of geometric origin, as $\theta$ necessarily vanishes as the tissue spreads fully with $R\to\infty$. The jump in $\theta$, which acts as the order parameter of the transition, is reminiscent of first-order phase transitions. However, unlike in those, our wetting transition features no metastable states, and hence no hysteresis \cite{Bonn2001}.

To clarify the nature of the transition, we rewrite our one-variable dynamical system $\dot{R} = V(R)$ in terms of an effective free energy: $\dot{R} = - \Gamma\, \dd F_\text{eff}/\dd R$, with $\Gamma$ a kinetic coefficient, such that $F_\text{eff} (R) = -\Gamma^{-1} \int V(R)\,\dd R$. The stable (unstable) points of the dynamics correspond to minima (maxima) of the free energy. As $\Ta$ increases, the minimum of $F_\text{eff}$ (filled circles in \cref{Fig effective_free_energy}) flattens and disappears at the transition point $\Ta^*$, beyond which the free energy decreases monotonically. Thus, at $\Ta^*$, the free energy develops an inflection point (\cref{Fig effective_free_energy}), which implies that the transition point is a critical point, as in second-order transitions. Therefore, as it combines features of first- and second-order transitions, the active wetting transition can be classified as a mixed-order transition \cite{Alert2017}.

\bigskip

\begin{figure}[tb!]
    \centering
    \includegraphics[width=\columnwidth]{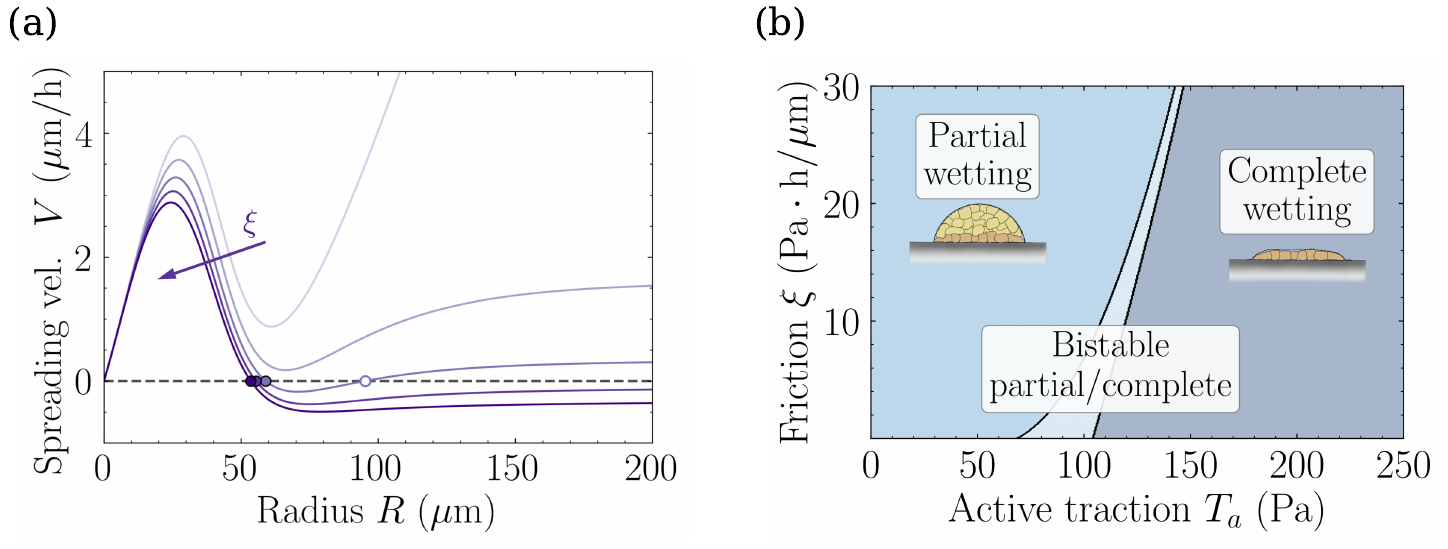}
  {\phantomsubcaption\label{Fig VR_friction}}
  {\phantomsubcaption\label{Fig phase_diagram_friction}}
    \bfcaption{Effect of substrate friction}{ (a) Increasing friction reduces the spreading velocity and makes it saturate for large contact radii $R$. Here, we set $\Ta = 115~\si{Pa}$, and we vary $\xi = 0, 5, 10, 15, 20~\si{Pa\cdot h/\mu m}$. Filled circles mark stable fixed points, corresponding to partial wetting, and empty circles mark unstable fixed points, above which there is complete wetting. (b) Phase diagram of active wetting in the presence of friction. Increasing friction promotes partial wetting and shrinks the region of bistability between partial and complete wetting. Other parameter values are listed in \cref{tab_app: parameters_2D}.}
    \label{fig:friction_2D}
\end{figure}

\noindent\textbf{Effect of substrate friction.} We now restore the friction term neglected above. The force balance \cref{eq force_balance} can still be solved analytically for a circular basal cell layer (\cref{solutions}). From the solution, we obtain the spreading velocity $V$ and analyze how it depends on the contact radius $R$ (\cref{Fig VR_friction}). Friction limits the spatial extent of the flow field in the cell layer. This is captured by the screening length $\lambda = \sqrt{\eta/\xi}$, which sets how far the flow driven by active tractions at the tissue edge penetrates into the cell cluster before being damped by friction. As friction increases, the flow field becomes limited to a peripheral region of width $\lambda$ instead of spanning the whole tissue. As a result, the spreading velocity ceases to increase with tissue size $R$ and instead saturates to an asymptotic value, which decreases with increasing friction (\cref{Fig VR_friction}). This reduction of the spreading velocity favors partial wetting, which occupies a larger region of the phase diagram as friction increases (\cref{Fig phase_diagram_friction}). Moreover, friction also narrows the bistable region in between the partial and complete wetting phases (\cref{Fig phase_diagram_friction}). Overall, friction limits the ability of active tractions to drive spreading, and hence it favors partial wetting.

Interestingly, increasing friction can trigger a transition from complete to partial wetting (\cref{Fig phase_diagram_friction}). This feature is characteristic of our active wetting transition. In passive wetting, a dynamical parameter like friction affects only the spreading dynamics, and not the static wetting state, which is set solely by surface tensions through the Young-Dupr\'{e} condition. The key distinction is that the active wetting states are not determined by a static balance of forces at the contact line, but by the full dynamic force balance (\cref{eq force_balance}), which involves flows across the whole tissue (\cref{solutions}).

\bigskip

\begin{figure}
    \centering
    \includegraphics[width=0.5\textwidth]{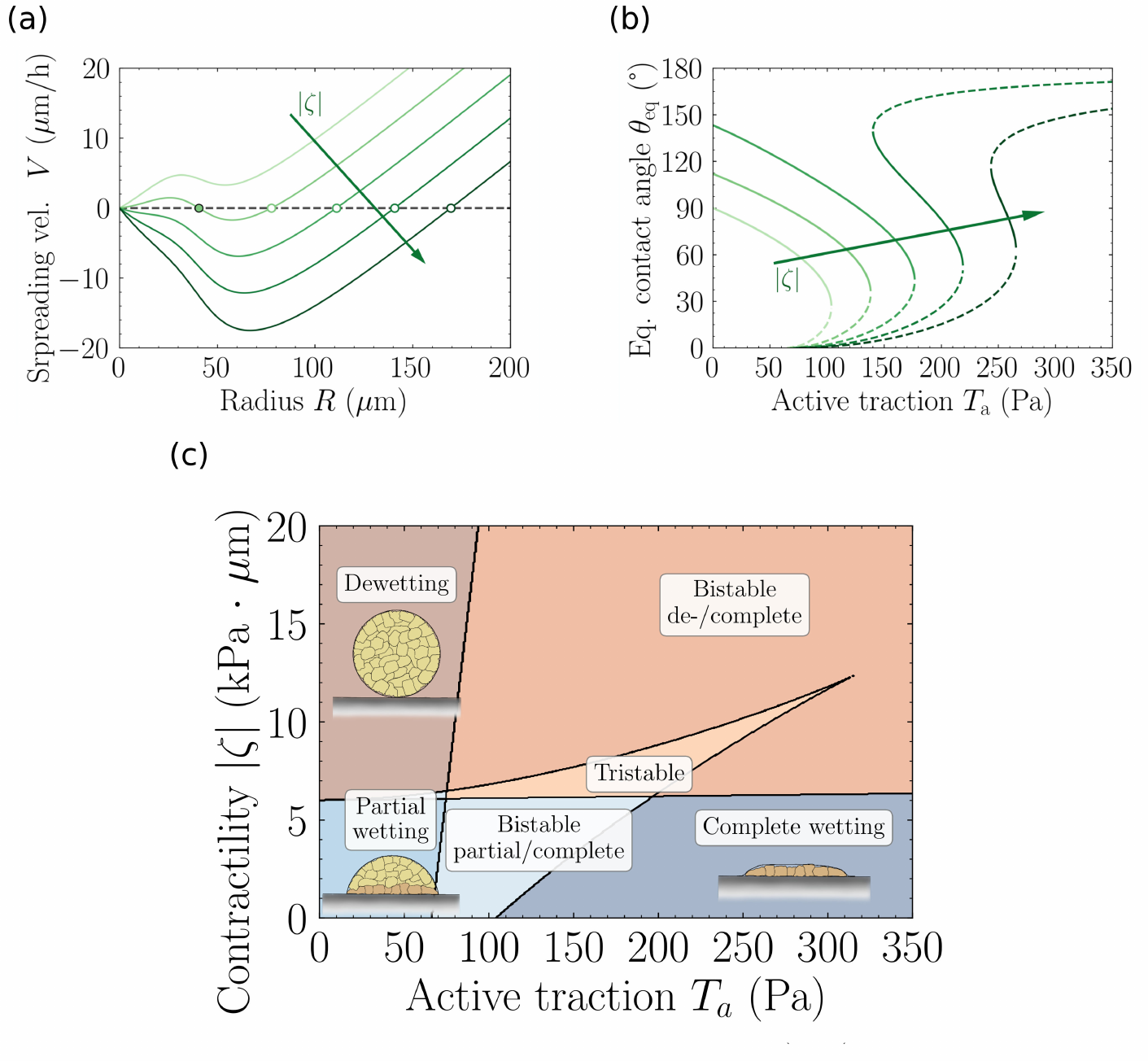}
  {\phantomsubcaption\label{Fig VR_contractility}}
  {\phantomsubcaption\label{Fig thetaeq_contractility}}
  {\phantomsubcaption\label{Fig phase_diagram_contractility}}
    \bfcaption{Effect of contractility}{ (a) Increasing contractility decreases the spreading velocity and thus favors first partial wetting and then full dewetting. When the velocity becomes negative at small $R$, small tissues retract and evolve towards full dewetting at $R=0$. Here, we set $\Ta = 115$ Pa, $\xi=0$, and we vary $\left|\zeta \right | = 0, 5, 10, 15, 20$ kPa$\cdot \mu$m. Filled circles mark stable fixed points, corresponding to partial wetting, and empty circles mark unstable fixed points, above which there is complete wetting. (b)~Stable (solid) and unstable (dashed) equilibrium contact angle as a function of $\Ta$ for increasing contractility. Contractility allows the equilibrium contact angle to exceed $90^\circ$. For high enough contractility, a new unstable branch appears at high contact angles, which separates partial wetting and full dewetting, which corresponds to $\teq = 180^\circ$. Here, we set $\xi=0$, and we vary $\left|\zeta\right| = 0, 2.5, 5, 7.5, 10$ kPa$\cdot \mu$m. (c)~Phase diagram in the $\Ta$--$\zeta$ plane, showing three pure phases (dewetting, partial wetting, and complete wetting), two bistable regions, and a tristable region in which all three phases are possible. In the multistable regions, the final state is selected by the initial contact radius $R(0)$ of the cluster. Other parameter values are listed in \cref{tab_app: parameters_2D}.}
    \label{fig:contractility_2D}
\end{figure}

\noindent\textbf{Effect of contractility.} We now reintroduce contractility, switched off so far, to analyze its effects. To this end, we solve the model analytically in the frictionless limit (\cref{solutions}). As shown in \cref{Fig theta_traction}, active traction and surface tension alone can only balance out for $\theta < 90^\circ$. Thus, their competition does not produce wetting states with $\theta >90^\circ$. Introducing contractility allows us to obtain such states. This is because contractile active stresses generate tension in the basal cell layer, which pulls the contact line back and thus competes with the outward-pointing active tractions. Therefore, contractility favors edge retraction, and can thus stabilize partial wetting states with $\theta > 90^\circ$ (\cref{Fig thetaeq_contractility}). When strong enough, contractility can also produce full dewetting, corresponding to $\Req =0$ and $\teq = 180^\circ$, consistent with Refs. \cite{Perez-Gonzalez2019,Nyga2021}.

As in previous cases, we obtain these results by analyzing how the spreading velocity depends on the contact radius, $V(R)$ (\cref{Fig VR_contractility}). Starting from a case of complete wetting without contractility (lightest green in \cref{Fig VR_contractility}), increasing contractility first allows the tissue to achieve partial wetting (filled circle in \cref{Fig VR_contractility}). Further increasing contractility, however, removes the partial wetting state, leaving only an unstable fixed point of the spreading dynamics (empty circles in \cref{Fig VR_contractility}). This unstable point corresponds to the wetting-dewetting transition identified in Ref.~\cite{Perez-Gonzalez2019}; it defines a critical radius that separates full dewetting ($\Req =0$, $\teq = 180^\circ$) from complete wetting ($\Req \to \infty$ and $\teq = 0$) states. Here, our results generalize those of Ref.~\cite{Perez-Gonzalez2019} by including the effects of tissue surface tension.

The appearance of a full dewetting state markedly enriches the phase behavior. The phase diagram (\cref{Fig phase_diagram_contractility}) now consists of three pure phases (full dewetting, partial wetting, and complete wetting) plus regions of bistability and tristability, in which the cluster reaches one of the possible states depending on its initial contact radius $R(0)$.

Finally, we analyze the role of friction in the presence of contractility. In this case, we can no longer solve the model analytically, and we solve the force balance equation numerically using finite differences (\cref{numerics}). As in the absence of contractility, friction limits the ability of tractions to accelerate the spreading velocity, which now reaches an asymptotic value (\cref{Fig VR_contractility_friction}). Hence, friction generally decreases the cluster's wettability. On the phase diagram, the dewetting and partial wetting regions expand, and the multistable regions shrink, while keeping their relative arrangement (\cref{Fig phase_diagram_contractility_friction}).

\bigskip

\noindent\textbf{Solution in one dimension.} Whereas we could not solve the complete model with friction and contractility analytically for a circular geometry, we solved it in one dimension. This case is relevant both for standard monolayer spreading experiments in a strip geometry as well as to understand situations in which the tissue migrates along one direction, such as in collective durotaxis \cite{Sunyer2016,Alert2019a,Pi-Jauma2022,Pallares2023}. Thus, for completeness, we provide the analysis of this case in \cref{1d}. The results are analogous to those in two dimensions.

\bigskip

\noindent\textbf{Discussion.} We presented an active-matter analogue of the wetting transition for cellular aggregates. Modeling the tissue as an active polar fluid droplet, we showed that the competition between surface tension and active tractions gives rise to a transition between partial and complete wetting, at which the contact angle jumps discontinuously. We found that, for a range of parameter values, both partial and complete wetting states can be reached depending on the initial contact radius of the cell aggregate. Adding active cell-cell contractile stresses to the model allows the tissue to fully dewet, and it opens a rich multistability landscape in which full dewetting, partial wetting, and complete wetting can all occur depending on tissue size. Previous work captured a transition between wetting and dewetting dynamics as a result of the competition between active traction and contractile stresses \cite{Perez-Gonzalez2019}. Here, by accounting for surface tension, our findings generalized those results and show how the interplay between active and capillary forces in cell aggregates can result in stable partial wetting states, which were not accounted for before.

The predicted partial wetting states could explain existing experimental observations. The contact angle of cell aggregates decreases with both substrate stiffness \cite{Douezan2012c,Pallares2023} and cell-substrate adhesion, tuned either by varying the substrate's protein coating \cite{Ravasio2015} or the cells' gene expression \cite{Conti2024}. Given that stiffer and more adhesive substrates enable stronger tractions \cite{Perez-Gonzalez2019,Pallares2023}, our results in \cref{Fig theta_traction} capture the experimental trends.

A distinctive feature of our results is that the wetting state depends on dynamical parameters of dissipative forces, such as viscosity and friction. In passive wetting, the equilibrium contact angle is determined solely by surface tensions through Young-Dupr\'{e}'s law, which imposes static force balance locally at the contact line \cite{DeGennes1985,DeGennes2004,Bonn2009}. In that picture, dissipative forces such as friction set only the rate of relaxation toward equilibrium, never the equilibrium itself. Here, instead, the steady-state contact angle arises from a dynamical and global force balance across the tissue between active, capillary, and dissipative forces. As a result, dynamical parameters contribute to selecting the wetting state. For example, we predicted that decreasing the hydrodynamic screening length, by either increasing the friction coefficient (\cref{Fig phase_diagram_friction}) or decreasing viscosity, can trigger a transition from complete to partial wetting. Recent experiments have shown that tissue viscosity can be tuned by proteins associated with cell-cell junctions in cancer cell aggregates \cite{Marchesi2026}, by the maturation of the keratin network in developing zebrafish \cite{Naik2026}, and by maternal age in mouse embryos \cite{Cavanaugh2026}, which suggest ways to test our predictions experimentally.

The selection of wetting states by dissipative, non-equilibrium effects might be a general hallmark of active wetting. In models of active Brownian particles, for instance, capillary rise and partial wetting are stabilized by self-organized particle currents around the contact line and the associated drag forces \cite{Zhao2026,Matsuzawa2026,Mangeat2024,FinsCarreira2024}. Similar currents also exist in models of chemically active wetting \cite{Liese2025}; there, activity arises from nonequilibrium binding of fluid molecules on the surface \cite{Zhao2021b,Zhao2024e}, which is relevant for biomolecular condensates wetting cell membranes \cite{Julicher2024}. In this context, our work puts forward a theory of active wetting for tissues. From a biological perspective, it provides a theoretical framework to rationalize morphological transitions in tissues during development and disease. From a physics perspective, it contributes to the ongoing effort to understand the different types of active wetting \cite{Thiele2026}.

\bigskip
\noindent\textbf{Acknowledgments.} R.A. acknowledges funding from the European Union through the ERC Starting Grant “Living\_Fluctuations” (No. 101114584).

\bibliography{active_wetting_abbreviated,preprints_abbreviated} 

\onecolumngrid
\clearpage
\twocolumngrid

\appendix

\setcounter{figure}{0}
\setcounter{table}{0}
\renewcommand{\thefigure}{\thesection\arabic{figure}}
\renewcommand{\thetable}{\thesection\arabic{table}}
\makeatletter
\@addtoreset{figure}{section}
\@addtoreset{table}{section}
\makeatother

\section{Analytical solutions in a circular geometry} \label{solutions}

Here, we provide the analytical solutions of our model for a circular basal cell layer in the different cases considered in the Main Text.

We assume radial symmetry, such that the polarity and velocity fields have the form $\mathbf{p} = p(r) \hat{\mathbf{r}}$ and $\mathbf{v} = v(r) \hat{\mathbf{r}}$. In this case, the force balance \cref{eq force_balance}, after introducing the constitutive relation \cref{eq stress}, reads \cite{Perez-Gonzalez2019}
\begin{equation}
    \eta\left[v'' + \frac{v'}{r} - \left(\frac{1}{r^2} + \frac{1}{\lambda^2}\right) v\right] = -\Ta\,p + \zeta \left(\frac{1}{r}p^2 + 2 pp' \right).
    \label{eq:SI_radial_formula}
\end{equation}
Here, the polarity profile $p(r)$, as indicated in the main text, is given by $p(r) = I_1(r/\Lc)/I_1(R/\Lc)$, and  $\lambda = \sqrt{\eta/\xi}$ is the hydrodynamic screening length set by  the balance between viscosity and substrate friction. We now present the analytical solutions of this equation for the velocity profile $v(r)$ in the different cases for which it is possible.

\bigskip
 
\begin{figure}[h!]
    \centering
    \includegraphics[width=0.9\linewidth]{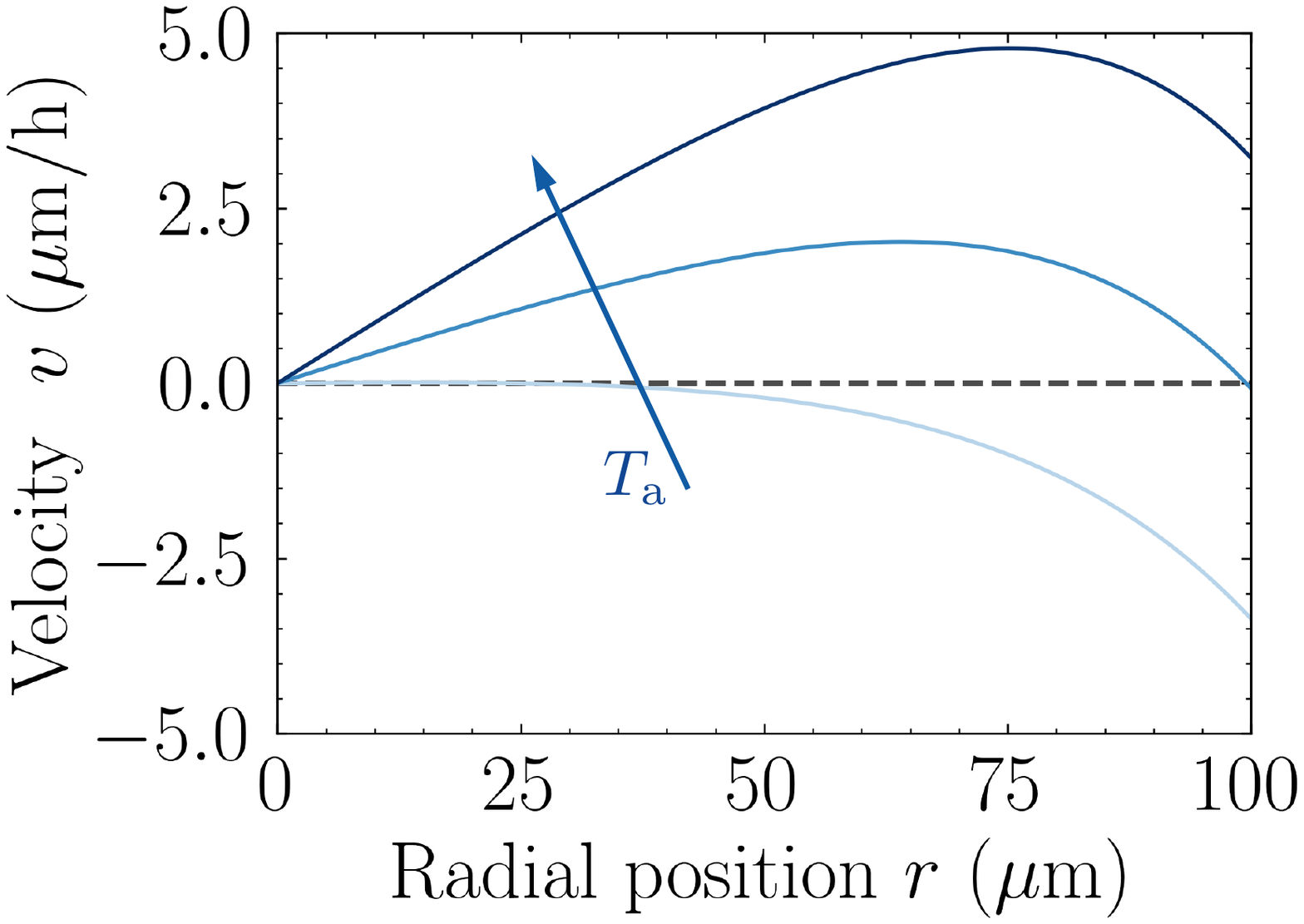}
    \bfcaption{Velocity field of the basal cell layer for varying active traction}{ Velocity field $v(r)$ of the basal cell layer for an aggregate of radius $R = 100 \,\si{\mu m}$, with $\zeta=0$ and $\xi=0$. The values of active traction are $\Ta = 70, 90, 110$ Pa; for the intermediate value, the velocity vanishes at the tissue edge, corresponding to a partial wetting state. Other parameter values are listed in \cref{tab_app: parameters_2D}.}
    \label{fig velocity_profile_activetraction}
\end{figure}

\noindent\textbf{Active traction and surface tension}. We first solve \cref{eq:SI_radial_formula} in the simplest setting, i.e., ignoring the effects of friction and contractility by setting $\xi=0$ and $\zeta=0$. Using both the boundary condition \cref{eq:Young-Dupre} for the stress at the edge of the basal monolayer and the symmetry condition on the velocity, $v(0) = 0$, the solution reads
\begin{multline}
    v(r) = \frac{1}{\eta} \Bigg[
        \left(-\gamma \cos \theta - \frac{\Ta\Lc^2}{R} \right) r
        \\
     + \Ta\Lc 
        \left(
            \frac{r I_0(R/\Lc) - \Lc I_1(r/\Lc)}
                 {I_1(R/\Lc)}
        \right)
    \Bigg],
\end{multline}
which we plot in \cref{fig velocity_profile_activetraction}.
The spreading velocity, \cref{eq:edge_velocity}, is then obtained by evaluating this expression at $r=R$, the radius of the basal layer: $V=v(R)$.

\bigskip

\begin{figure}[tbh!]
    \centering
    \includegraphics[width=0.9\linewidth]{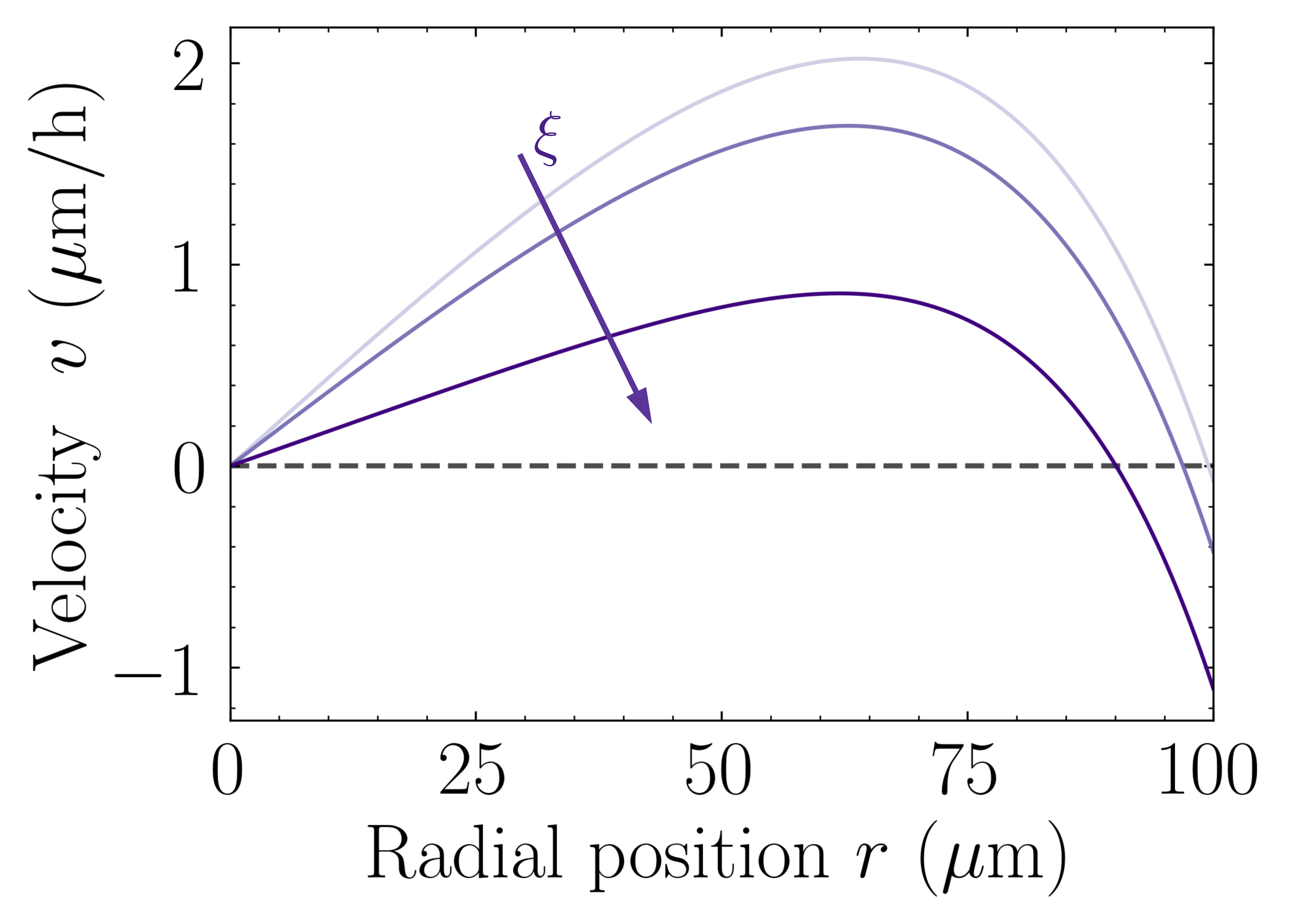}
    \caption{\textbf{Velocity field of the basal cell layer for varying friction}. Velocity field $v(r)$ of the basal cell layer for an aggregate of radius $R = 100 \,\si{\mu m}$, with $\zeta = 0$. The values of friction are $\xi = 0, 1, 10$ $\si{Pa \cdot h / \mu m}$. The value of $\Ta = 90$ Pa corresponds to the vanishing velocity at the tissue edge in \cref{fig velocity_profile_activetraction}. Other parameter values are listed in \cref{tab_app: parameters_2D}.}
    \label{fig velocity_profile_friction}
\end{figure}

\noindent\textbf{Active traction, surface tension, and friction}. Restoring the friction term in \cref{eq:SI_radial_formula}, still without contractility, together with the boundary condition \cref{eq:Young-Dupre}, the solution for the velocity profile reads
\begin{align}
    v(r) &= -\frac{T_* \Lc^2}{\eta}\,\frac{I_1(r/\Lc)}{I_1(R/\Lc)} 
    + \mathcal{A}\, I_1(r/\lambda),
\end{align}
with the effective screened-traction amplitude
\begin{equation}
    T_* = \Ta\frac{\lambda^2}{\lambda^2 - \Lc^2},
\end{equation}
and the homogeneous-term coefficient
\begin{equation}
    \mathcal{A} = \frac{T_*\Lc^2\,(\mathcal{R}_{\Lc} - 1) - \gamma R \cos\theta}
                       {\eta\, I_1(R/\lambda)\,(\mathcal{R}_\lambda - 1)},
\end{equation}
where we have introduced the shorthand
\begin{equation}
    \mathcal{R}_x \equiv \frac{R\, I_0(R/x)}{x\, I_1(R/x)},
    \qquad x \in \{\Lc,\lambda\}.
    \label{eq:SI_Rx}
\end{equation} 
We plot the resulting velocity profiles for increasing friction coefficient in \cref{fig velocity_profile_friction}. Evaluating at $r=R$ yields the spreading velocity
\begin{align}
    V &= \frac{T_* \Lc^2\,(\mathcal{R}_{\Lc} - \mathcal{R}_\lambda) 
                    - \gamma R \cos\theta}
                  {\eta\,(\mathcal{R}_\lambda - 1)}.
\end{align}
In the frictionless limit $\xi \to 0$ ($\lambda \to \infty$), we have $T_* \to \Ta$ 
and $\mathcal{R}_\lambda \to 2$, and the expression above reduces to the 
unscreened edge velocity given in \cref{eq:edge_velocity}.

\newpage

\begin{figure}[h!]
    \centering
    \includegraphics[width=0.9\linewidth]{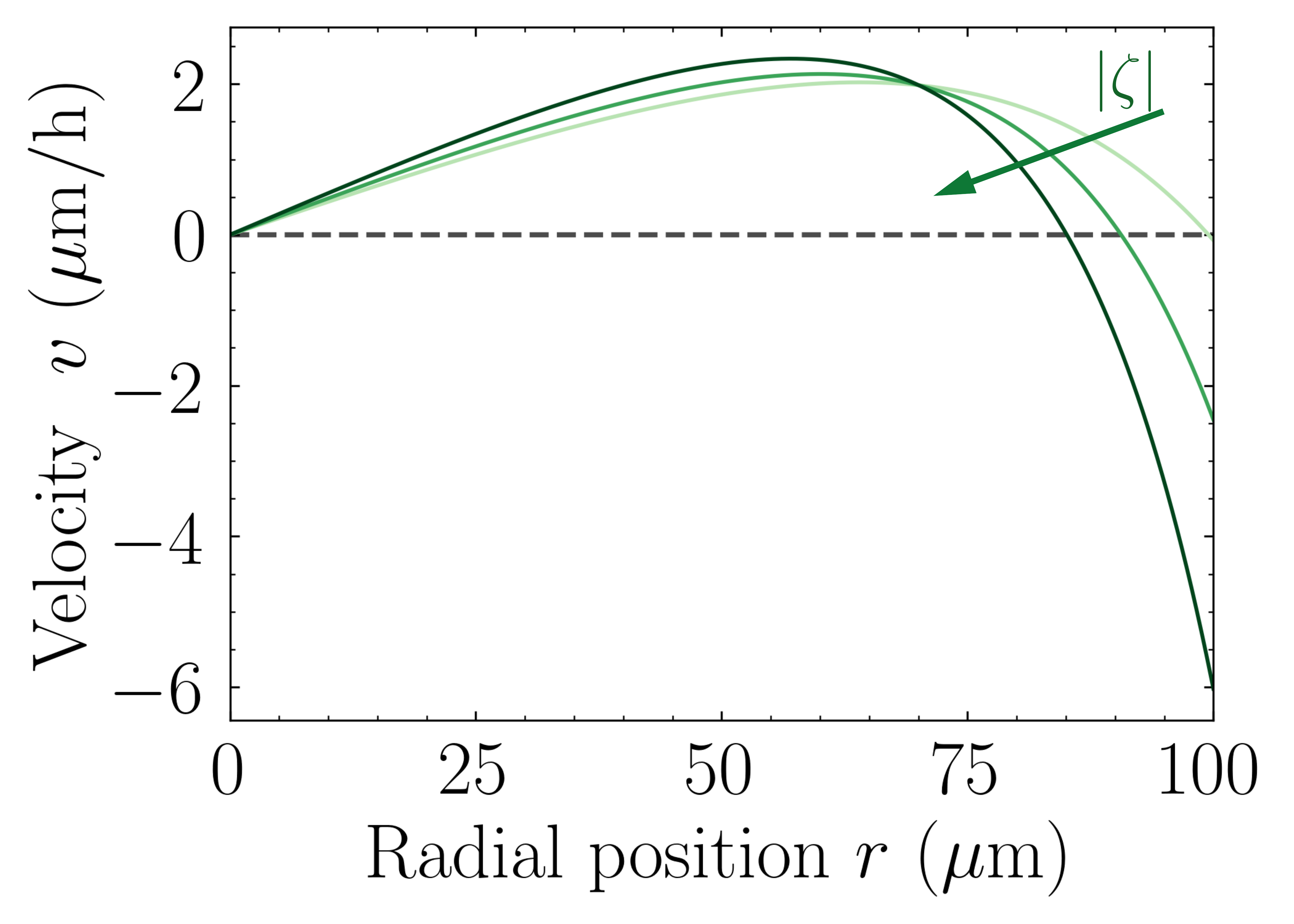}
    \caption{\textbf{Velocity field of the basal cell layer for varying contractility}. Velocity field $v(r)$ of the basal cell layer for an aggregate of radius $R = 100 \,\si{\mu m}$, with $\xi = 0$. The values of contractility are $|\zeta| = 0, 2, 10$ $\si{kPa \cdot \mu m}$. The value of $\Ta = 90$ Pa corresponds to the vanishing velocity at the tissue edge in \cref{fig velocity_profile_activetraction}. Other parameter values are listed in \cref{tab_app: parameters_2D}.}
    \label{fig velocity_profile_contractility}
\end{figure}

\noindent\textbf{Active traction, surface tension, and contractility}. We now restore 
the contractility term while still neglecting friction. With the same boundary condition \cref{eq:Young-Dupre} and symmetry 
condition $v(0) = 0$, the solution reads
\begin{equation}
    v(r) = \frac{1}{2\eta}\Big[\,\mathcal{C}_1\, r
        + \mathcal{C}_2(r)\, I_1(r/\Lc)\,\Big],
\end{equation}
with
\begin{multline}
    \mathcal{C}_1 = \zeta - 2\gamma\cos\theta - \frac{2\Ta \Lc^2}{R} \\
    + \left(\frac{\zeta \Lc}{R} + 2\Ta \Lc\right)
                \frac{I_0(R/\Lc)}{I_1(R/\Lc)}  - \zeta\,\frac{I_0^2(R/\Lc)}{I_1^2(R/\Lc)},
\end{multline}
\begin{equation}
    \mathcal{C}_2(r) = \frac{\Lc}{I_1(R/\Lc)}
        \left[\frac{\zeta\, I_0(r/\Lc)}{I_1(R/\Lc)} - 2\Ta \Lc\right].
\end{equation}
We plot the resulting velocity profile for increasing contractility in \cref{fig velocity_profile_contractility}. Using $\mathcal{R}_{\Lc}$ from \cref{eq:SI_Rx}, the spreading velocity obtained as $V=v(R)$ reads
\begin{multline}
    V = \frac{1}{2\eta}\Bigg\{
        \frac{\zeta}{R}\Big[R^2 + \Lc^2\,\mathcal{R}_{\Lc}(2 - \mathcal{R}_{\Lc})\Big]\\
    \quad + 2\Ta \Lc^2\,(\mathcal{R}_{\Lc} - 2) - 2\gamma R \cos\theta
    \Bigg\}.
\end{multline}

\section{Numerical solution of the full model} \label{numerics}

\begin{figure}[h!]
    \centering
    \includegraphics[width=\linewidth]{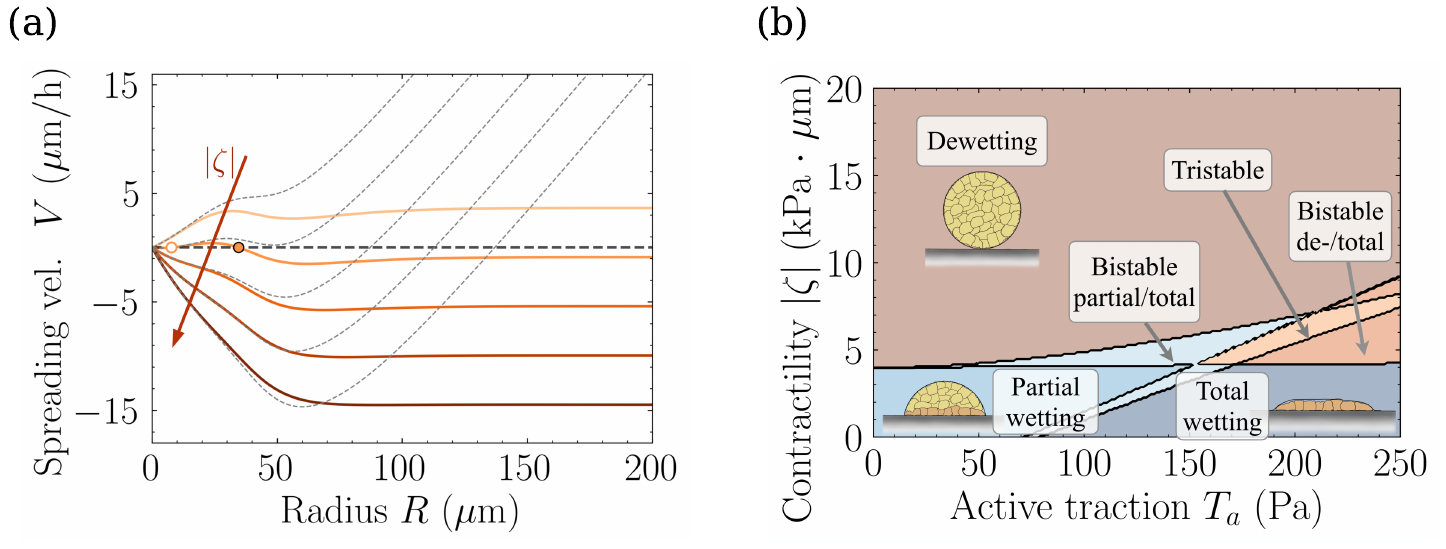}
  {\phantomsubcaption\label{Fig VR_contractility_friction}}
  {\phantomsubcaption\label{Fig phase_diagram_contractility_friction}}
    \bfcaption{Effect of contractility and friction}{ (a)~Friction decreases the spreading velocity and makes it saturate at for large contact radii $R$. Here, we set $\gamma = 1$ mN/m, $\Ta = 150$ Pa, and $\xi = 10 \ \rm{Pa}\cdot \rm{h}/\mu\rm{m}$ and we vary contractility $|\zeta| = 0, 5, 10, 15, 20$ $\si{kPa \cdot \mu m}$. The filled circle marks a stable fixed point, corresponding to partial wetting. Dashed lines show the frictionless solution with varying contractility, as in \cref{Fig VR_contractility}. (b)~Phase diagram in the $\Ta$--$\zeta$ plane. Compared to the frictionless case in \cref{Fig phase_diagram_contractility}, the full dewetting and partial wetting regions expand, and the multistable regions shrink. Here, $\gamma = 1.5$ mN/m. Other parameter values are listed in \cref{tab_app: parameters_2D}.}
    \label{fig:phase_map_numeric}
\end{figure}

We could not analytically solve the full model, including both friction and contractility. Therefore, we solve it numerically using the standard finite-differences method. \Cref{fig:phase_map_numeric} shows the numerical solutions for the spreading velocity, compared with those of the frictionless case. We also show the resulting phase diagram in the $\Ta$--$\zeta$ plane at non-zero friction $\xi$, to be compared with the frictionless case of \cref{Fig phase_diagram_contractility}. The role of friction is the same as in the absence of contractility: through the screening length $\lambda=\sqrt{\eta/\xi}$, friction damps the flow and hinders multistability. In particular, the tristable region shrinks and ultimately disappears as $\xi$ increases, while the boundaries between the pure dewetting, partial-wetting, and complete-wetting phases are only quantitatively displaced. The qualitative organization of the phase diagram is therefore preserved, confirming that friction hinders spreading and hence promotes dewetting and partial wetting in front of complete wetting.

\begin{table}[h]
    \centering
    \begin{tabular}{lll}
        \toprule
        \textbf{Parameter} & \textbf{Description} & \textbf{Typical value} \\
        \midrule
        $\Ta$ & Active traction magnitude & $100\,\si{Pa}$ \\
        $\eta$ & Monolayer viscosity & $10^4\,\si{Pa}\cdot\si{h}\cdot\si{\um}$ \\
        $\xi$ & Friction coefficient & $10\,\si{Pa}\cdot\si{h}/\si{\um}$ \\
        $\lambda = \sqrt{\eta/\xi}$ & Hydrodynamic screening length & $30\,\si{\um}$ \\
        $-\zeta$ & Contractility & $10^3 \,\si{Pa \cdot \um}$ \\
        $\Lc$ & Polarity decay length & $25\,\si{\um}$ \\
        $\gamma$ & Surface tension & $1.5\,\si{mN/m}$ \\
        $\mathcal{V}$ & Droplet volume & $10^{5}\,\si{\um^3}$ \\
        \bottomrule
    \end{tabular}
    \bfcaption{Parameter estimates used in the 2D model}{}\label{tab_app: parameters_2D}
\end{table}



\section{Analytical solutions in a one-dimensional geometry} \label{1d}

Here, we provide the analytical solution of the full model in a one-dimensional geometry.
This corresponds to an infinitely extended strip in the $y$-direction. 
Unlike in the circular geometry considered above, in one dimension (1d) we can analytically solve the full model including both friction and contractility. We then analyze the wetting transition in this simpler setting.

\bigskip

\noindent\textbf{Spreading velocity.}
In 1d, for a tissue that spreads along the $x$ axis, the force balance \cref{eq force_balance} together with the constitutive relation \cref{eq stress} reads
\begin{equation} \label{eq: force balance 1d}
    2 \eta\, \partial_x^2 v(x) - \xi \, v(x)=  - \Ta \, p(x) + \zeta\, p(x) \partial_x p(x).
\end{equation}
In 1d, the solution to \cref{eq polarity} for the polarity field is given by 
\begin{equation} \label{eq: polarity_field_1d}
    p(x) = \frac{\sinh{(x/\Lc)}}{\sinh{(L/\Lc)}},
\end{equation}
where $L$ is the semi-width of the tissue strip, which spans from $x=-L$ to $x=L$, and it plays the role of the contact radius $R$ in the 2d circular setup. The Young-Dupr\'{e} boundary condition \cref{eq:Young-Dupre} in 1d reads
\begin{equation}
    \sigma (x = \pm L)= 2 \eta \, \partial_x v |_{x=\pm L} - \zeta = - \gamma \cos\theta,
\end{equation}
with $\theta$ being the contact angle of the cluster (\cref{fig:schematic_model}).

The solution of \cref{eq: force balance 1d} for the velocity field is given by
\begin{multline}
    v(x) = C_1 \sinh{\left(\frac{x}{\lambda} \right)} + C_2 \cosh{\left(\frac{x}{\lambda} \right)}\\
    + \frac{\zeta}{\xi}\frac{\Lc}{4\lambda^2 - \Lc^2}\frac{\sinh{(2 x/\Lc)}}{\sinh^2{(L/\Lc)}}-\frac{\Ta}{\xi} \frac{\Lc^2}{\lambda^2 - \Lc^2} \frac{\sinh{(x/\Lc)}}{\sinh{(L/\Lc)}}, 
\end{multline}
where the integration constants are
\begin{multline}
    C_1 = \frac{\lambda}{\cosh{\left(\frac{L}{\lambda} \right)}} \left[- \frac{2\zeta}{\xi}\frac{1}{4\lambda^2 - \Lc^2}\frac{\cosh{(2 L/\Lc)}}{\sinh^2{(L/\Lc)}}\right.\\
    +\left. \frac{\Ta}{\xi} \frac{\Lc}{\lambda^2 - \Lc^2} \frac{1}{\tanh{(L/\Lc)}} - \frac{\gamma}{2 \eta} \cos \theta + \frac{\zeta}{2 \eta}\right],
\end{multline}
\begin{equation}
    C_2 = 0.
\end{equation}
In all these expressions, we have redefined the hydrodynamic screening length to $\lambda = \sqrt{2 \eta / \xi}$, where, for convenience, we have added a factor of two as compared to our solutions in the circular geometry.

\begin{figure}[ht]
    \centering
    \includegraphics[width=\columnwidth]{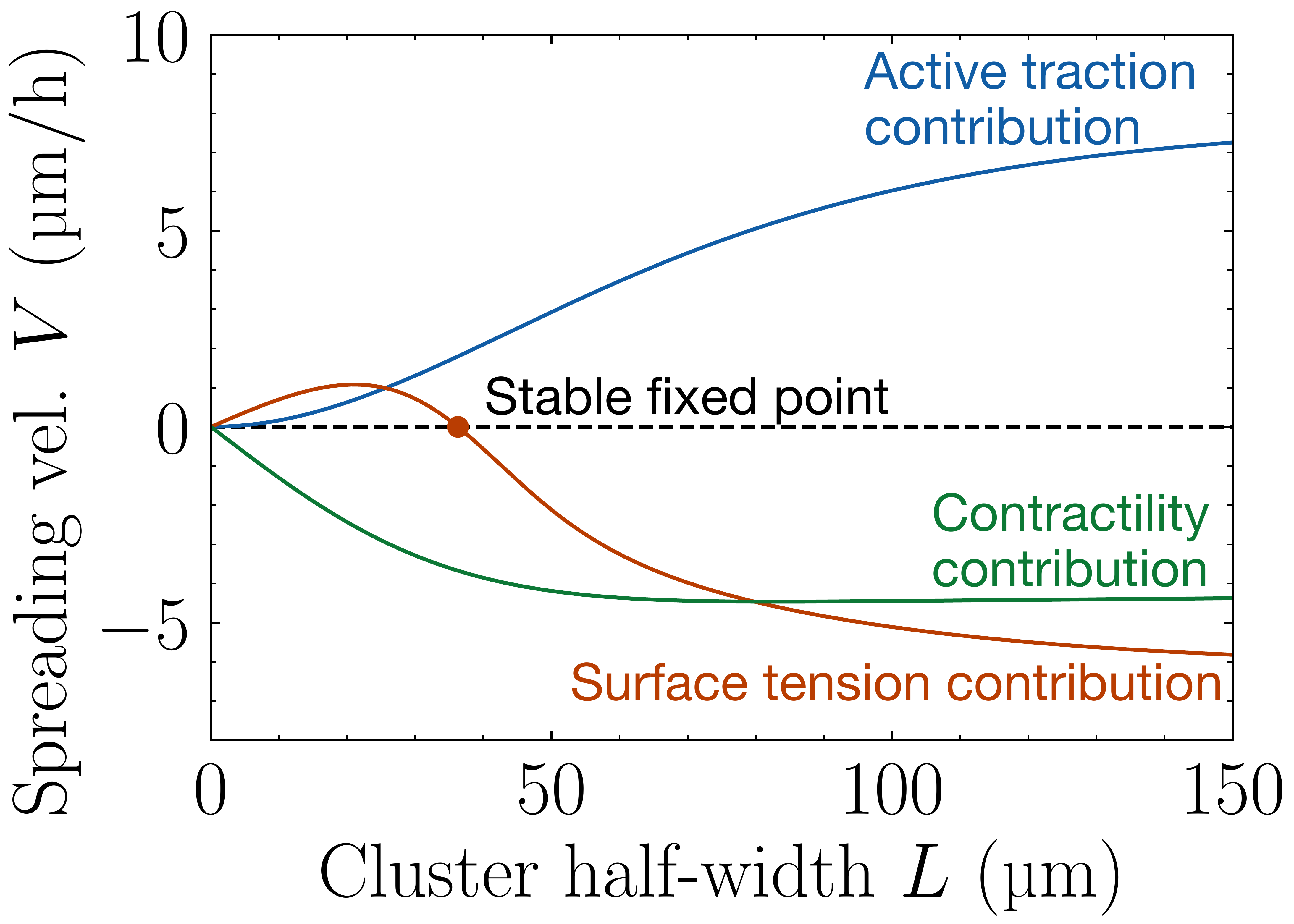}
     \bfcaption{Contributions to tissue wetting}{ 
     Active traction pulls outward on the contact line and promotes complete wetting (blue). Contractility pulls inwards and promotes full dewetting (green). Surface tension pulls outwards at small cluster half-widths, for $\theta>90^\circ$, and inwards at large half-widths, for $\theta>90^\circ$ (red). Combined with the other contributions, it enables the emergence of a partial wetting state. Blue curve: $\Ta = 100 \,\si{Pa}$, $\gamma = 0.001 \, \si{mN/m}$, $\zeta = 0.001 \,\si{Pa \cdot \um}$; red curve: $\Ta = 0.001 \,\si{Pa}$, $\gamma = 1.5 \, \si{mN/m}$, $\zeta = 0.001 \,\si{Pa \cdot \um}$; green curve: $\Ta = 0.001 \,\si{Pa}$, $\gamma = 0.001 \, \si{mN/m}$, $\zeta = -8 \,\si{kPa \cdot \um}$. The rest of parameter values are listed in \cref{tab_app: parameters_1d}.}
    \label{fig_app: 1d_pedagogical}
\end{figure}

Finally, the spreading velocity, defined as $V = v(L)$, is given by
\begin{multline}
    V = \frac{\Ta}{\xi} \left[ \frac{\lambda \Lc}{\lambda^2-\Lc^2} \frac{\tanh{\left(\frac{L}{\lambda}\right)}}{\tanh{\left(\frac{L}{\Lc}\right)}} - \frac{\Lc^2} {\lambda^2 - \Lc^2}\right]  \\
     - \frac{\gamma}{\xi} \cos{\theta}  \frac{\tanh{\left(\frac{L}{\lambda}\right)}}{\lambda} \\
     + \frac{\zeta}{\xi} \left[ \frac{\tanh\!\left(\frac{L}{\lambda}\right)}{\lambda} + \frac{\Lc}{4\lambda^2- \Lc^2} \frac{\sinh\!\left(\frac{2L}{\Lc}\right)}{\sinh^2\!\left(\frac{L}{\Lc}\right)}  \right. \\
    \left. - \frac{2 \lambda}{4\lambda^2-  \Lc^2}\frac{\cosh\!\left(\frac{2 L}{\Lc}\right)}{\sinh^2\!\left(\frac{L}{\Lc}\right)}  \tanh\!\left(\frac{L}{\lambda}\right)  \right].
\end{multline}
We have grouped the terms by their corresponding driving forces. We plot each of these terms separately in \cref{fig_app: 1d_pedagogical}. As discussed in the Main Text, the active traction $\Ta$ term promotes complete wetting (blue), the contractility $\zeta$ term promotes dewetting (green) and the surface tension $\gamma \, \cos{\theta}$  promotes spreading for $\theta>90^\circ$  and retraction for $\theta<90^\circ$ (red).

\bigskip

\begin{figure}[tbh!]
    \centering
    \includegraphics[width=\columnwidth]{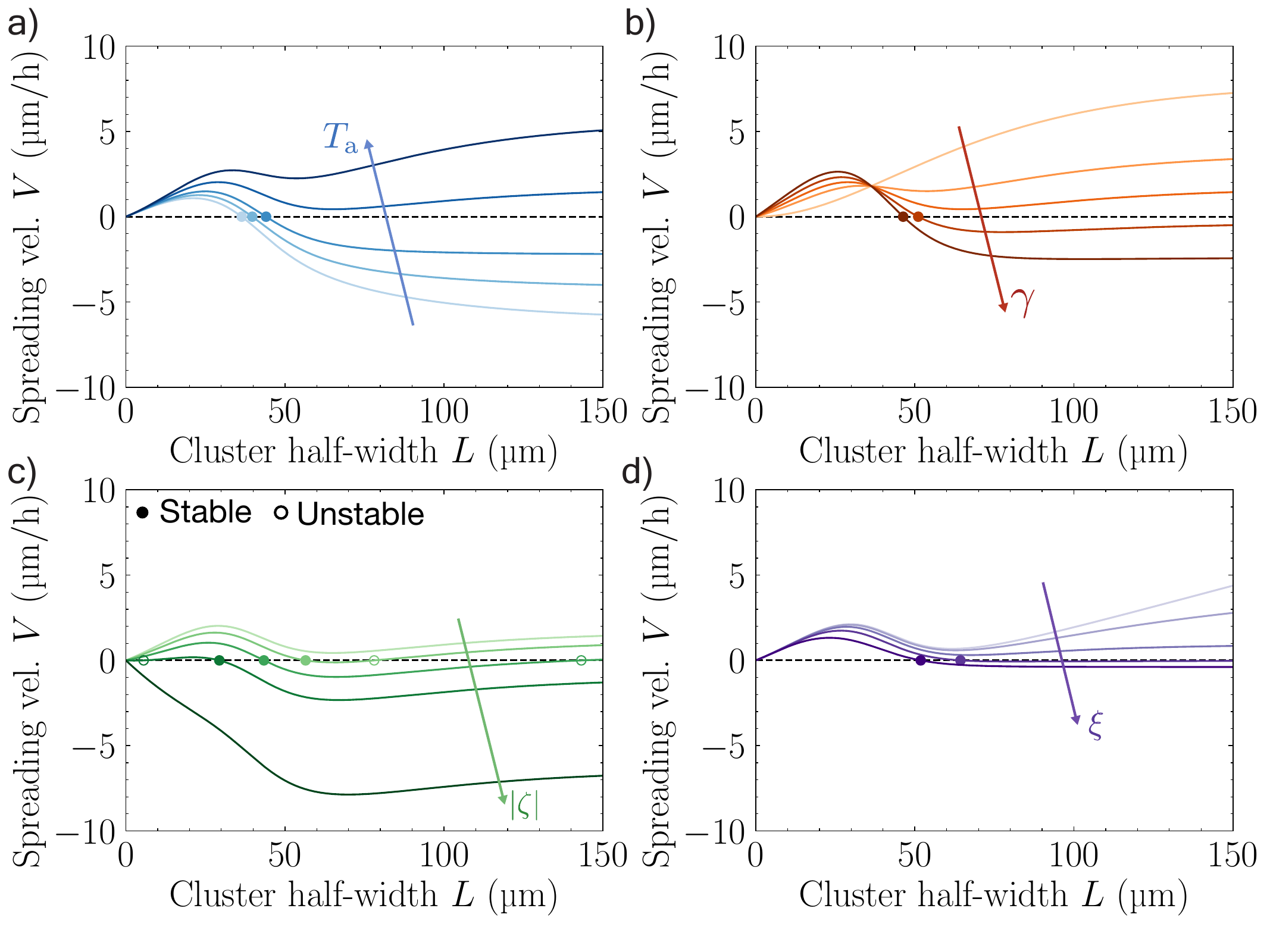}
    {\phantomsubcaption\label{fig_app: sprV_Ta}}
    {\phantomsubcaption\label{fig_app: sprV_gamma}}
    {\phantomsubcaption\label{fig_app: sprV_zeta}}
    {\phantomsubcaption\label{fig_app: sprV_xi}}
     \bfcaption{Active wetting transition in 1d}{ 
     Spreading velocity as a function of contact length L for increasing values of four different parameters: 
     (a) $\Ta = 1,\, 25,\, 50,\, 100,\, 150\, \si{Pa}$, 
     (b) $\gamma = 0.001,\, 1,\, 1.5,\, 2,\, 2.5\, \si{mN/m}$, 
     (c) $\zeta =0,\,-1,\,-2.55,\,-5, \, -15 \,  \si{kPa} \cdot \si{\um}$ and 
     (d) $\xi = 10^{-6},\, 1,\, 5,\, 15,\, 50\, \si{Pa}\cdot \si{h/\um}$. 
     We set contractility $\zeta=0$ in all panels except (c). All other parameters not varied in each plot are listed in \cref{tab_app: parameters_1d}.}
    \label{fig_app: 1d_spreading_V}
\end{figure}

\noindent\textbf{Active partial-to-complete wetting transition.}
We now repeat the analysis of the wetting transition in the Main Text but for the 1d solutions. 
We analyze the transition by varying one parameter at a time.
Increasing the active traction coefficient $\Ta$ increases the equilibrium length $\Leq$ given by the stable fixed point of the spreading dynamics: $V(\Leq)=0$ (\cref{fig_app: sprV_Ta}). Above a critical $\Ta$, the stable fixed point vanishes and the system transitions to complete wetting. On the other hand, increasing the surface tension $\gamma$ induces a transition from complete wetting (no fixed point) to partial wetting, marked by the emergence of a stable fixed point that sets the equilibrium contact length $\Leq$ (\cref{fig_app: sprV_gamma}).

\begin{figure}[tbh]
    \centering
    \includegraphics[width=\columnwidth]{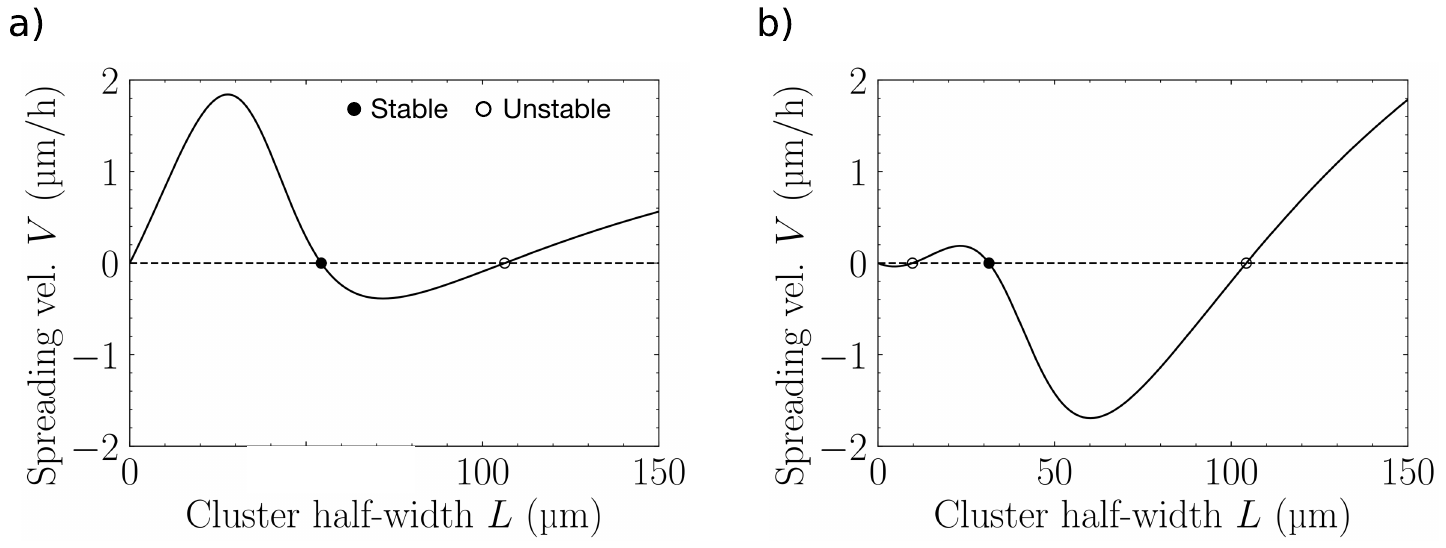}
    {\phantomsubcaption\label{fig_app: bistable}}
    {\phantomsubcaption\label{fig_app: tristable}}
     \bfcaption{Examples of bistable and tristable states}{ 
     (a) Bistable state with $\zeta = 0 \, \si{kPa} \cdot \si{\um}$, $\xi = 1 \, \si{Pa}\cdot \si{h/\um}$,$\Ta =80\, \si{Pa}$, and $\gamma = 1.5 \, \si{mN/m} $. For sizes $L< L_\text{u}$ below the unstable fixed point, the cluster evolves towards the partial wetting state given by the stable fixed point. For $L>L\text{u}$, it spreads to complete wetting.
     (b) Introducing contractility enables tristability. 
     Here, parameter values are $\zeta = -5.5 \, \si{kPa} \cdot \si{\um}$, $\xi = 1 \, \si{Pa}\cdot \si{h/\um}$, $\Ta =120\, \si{Pa}$, and $\gamma = 1.5 \, \si{mN/m} $.
     All other parameters are listed in \cref{tab_app: parameters_1d}. Below the small-$L$ unstable fixed point, the cluster evolves towards full dewetting ($L\to 0$). For sizes between the two unstable fixed points, the system achieves partial wetting given by the stable fixed point. Above the large-$L$ unstable fixed point, the cluster spreads to complete wetting.}
    \label{fig_app: 1d_bistable_tristable}
\end{figure}

To capture this transition, we take the limit of large cluster sizes ($L\gg \Lc,\lambda$) while neglecting contractility ($\zeta = 0$). In this limit, we have
\begin{align} \label{eq largeL}
    V \approx \frac{1}{\xi \lambda}
    \left(\Ta\, \frac{\lambda \Lc }{\lambda + \Lc}  - \gamma\right).
\end{align}
At small contact lengths, both surface tension and active traction pull outward, meaning that $V>0$ at small $L$ (in the absence of contractility, which could reverse this behavior). Therefore, if $V<0$ at large $L$, there must be a fixed point at which $V(\Leq)=0$, and hence there is a stable partial wetting state. From \cref{eq largeL}, the condition that $V<0$ at large $L$ implies
\begin{equation}
    \frac{\gamma}{\Ta} > \frac{\lambda \Lc}{\lambda + \Lc}.
\end{equation}
This is a sufficient condition for the existence of partial wetting. It is not a necessary condition because partial wetting can also exist if the spreading velocity turns positive again at large $L$, as shown in \cref{fig_app: bistable}, which introduces an unstable fixed point at $L_\text{u}>\Leq$. 

\bigskip

\noindent\textbf{Effect of contractility. Dewetting transition.}
When including contractility and increasing its magnitude, the system transitions from a complete wetting state to a bistable state and finally to full dewetting (\cref{fig_app: sprV_zeta}).
Interestingly, a new unstable fixed point emerges with $L_\text{u} < \Leq$ for large contractility (\cref{fig_app: tristable}). This new fixed point gives rise, for example, to tristability between dewetting, partial, and complete wetting, as shown in \cref{fig_app: tristable}.

The emergence of the dewetting state can be understood by studying the contributions to the spreading velocity at small contact lengths. For the full dewetting state to exist, the velocity must achieve negative values as $L\to 0$ (which implies $\theta\to 180^\circ$).
A small-$L$ expansion of the spreading velocity gives
\begin{equation}
    V \simeq \left(\gamma  + \frac{\zeta}{3} \right)\frac{L}{\xi \lambda^2} +  \frac{\Ta}{\xi}\, \frac{L^2}{3\lambda^2}.
\end{equation}
For small contact lengths $L$, the velocity is negative if $|\zeta| > 3 \gamma$. This is therefore a sufficient (but not necessary) condition for the existence of full dewetting. 

\bigskip

\begin{figure}[tb!]
    \centering
    \includegraphics[width=\columnwidth]{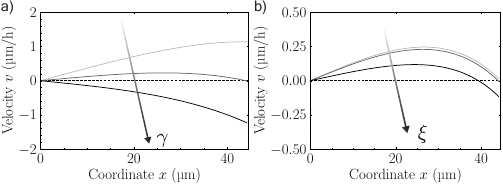}
    {\phantomsubcaption\label{fig_app: velocity_field_gamma}}
    {\phantomsubcaption\label{fig_app: velocity_field_xi}}
     \bfcaption{Velocity field of the basal cell layer}{ 
     Velocity field $v(x)$ of the basal cell layer for a cluster of length $L=\Leq=44.1\,\si{\um}$. 
     The center of the cluster is denoted by $x=0$.
     Parameters are chosen such that the medium curve in both figures is at a partial wetting steady state, with vanishing edge velocity $V=0$, which corresponds to the medium blue curve in \cref{fig_app: sprV_Ta} with $\Ta = 50\,\si{Pa}$. 
     (a) Varying surface tension $\gamma = 0.1, 1.5, 3\, \si{mN/m}$. 
     (b) Varying friction $\xi = 0.5,\, 3,\, 50\, \si{Pa}\cdot \si{h/\um}$. 
     All other parameter values are listed in \cref{tab_app: parameters_1d}.}
    \label{fig_app: 1d_velocity_field}
\end{figure}

\noindent\textbf{Friction can induce the wetting transition.}
Increasing friction reduces the spreading velocity and can thereby promote the emergence of partial wetting (\cref{fig_app: sprV_xi}).
To understand how friction induces the wetting transition, we analyze the velocity profile $v(x)$ in the basal layer for a tissue in the partial wetting state (\cref{fig_app: 1d_velocity_field}).

We first consider changes in surface tension. The parameters for the medium gray curve in \cref{fig_app: velocity_field_gamma} and \cref{fig_app: velocity_field_xi} were chosen to correspond to the medium blue curve in \cref{fig_app: sprV_Ta}, which has a partial wetting state with an equilibrium length $\Leq=44.1\,\si{\um}$.
In \cref{fig_app: velocity_field_gamma}, by varying the surface tension to lower (light gray) or larger (black) values, the edge velocity goes from positive to negative. 
In addition, the magnitude and sense of the flows in the cluster changes, revealing that the surface tension, which enters as a boundary condition, alters the flows within the tissue. 
Due to the active forces generating flows within the basal layer and the hydrodynamic length transmitting these flows, a droplet at partial wetting, which has a steady-state length with vanishing spreading velocity, has persistent flows. Due to incompressibility of the cell aggregate, these flows in the basal cell layer induce flows throughout the aggregate.
Persistent inner flows were recently reported in confined cell aggregates \cite{Yousafzai2024}, which neither spread nor retract, similar to those in our partial wetting states.

We now vary the friction coefficient $\xi$ in \cref{fig_app: velocity_field_xi}.
An increased friction coefficient (black) dampens the magnitude of the flows everywhere and causes a negative spreading velocity at the edge, thus leading to retraction towards a state with higher contact angle. 
Decreasing the friction coefficient (light gray) causes a positive spreading velocity and therefore promotes spreading into a state with lower contact angle. 

\begin{table}[tbh!]
    \centering
    \begin{tabular}{lll}
        \toprule
        \textbf{Parameter} & \textbf{Description} & \textbf{Typical value} \\
        \midrule
        $\Ta$ & Active traction magnitude & $100\,\si{Pa}$ \\
        $\eta$ & Monolayer viscosity & $10^4\,\si{Pa}\cdot\si{h}\cdot\si{\um}$ \\
        $\xi$ & Friction coefficient & $3\,\si{Pa}\cdot\si{h}/\si{\um}$ \\
        $\lambda = \sqrt{2\eta/\xi}$ & Hydrodynamic screening length & $81.6\,\si{\um}$ \\
        $\Lc$ & Polarity decay length & $25\,\si{\um}$ \\
        $\gamma$ & Surface tension & $1.5\,\si{mN/m}$ \\
        $\mathcal{V}$ & Droplet volume & $10^{5}\,\si{\um^3}$ \\
        \bottomrule
    \end{tabular}
    \bfcaption{Parameter estimates used in the 1d model}{}\label{tab_app: parameters_1d}
\end{table}

\end{document}